\documentclass[aps,prd,showpacs,nofootinbib,preprintnumbers,floatfix,twocolumn]{revtex4}

\usepackage{graphics}
\usepackage{graphicx}
\usepackage{tabularx}
\usepackage{makeidx}
\usepackage{epsfig}
\usepackage[caption=false]{subfig}
\usepackage{colortbl}
\usepackage{colordvi}
\usepackage{verbatim}
\usepackage{amsmath, amsthm}
\usepackage{enumerate}
\usepackage{hyperref}
\usepackage{dcolumn}
\usepackage{bm}
\usepackage{amssymb,mathrsfs}

\newcommand{\beq}{\begin{equation}} 
\newcommand{\eeq}{\end{equation}} 
\newcommand{\bea}{\begin{eqnarray}} 
\newcommand{\eea}{\end{eqnarray}} 
\def\benu{\begin{enumerate}}
\def\eenu{\end{enumerate}}
\def\nn{\nonumber}
\def\nn{\nonumber}

\def\pa{{\partial}}
\def\l{\left}
\def\r{\right}
\def\d{{\rm d}}
\def\f{\frac}
\def\tx{\tilde x}

\newcommand{\viz}{\textit{viz.~}}

\begin{document}
 
\title{Response of a rotating detector coupled to a polymer quantized field}
\author{D. Jaffino Stargen}
\email{jaffino@physics.iitm.ac.in}
\author{Nirmalya Kajuri}
\email{nirmalya@physics.iitm.ac.in}
\author{L. Sriramkumar}
\email{sriram@physics.iitm.ac.in}
\affiliation{Department of Physics, Indian Institute of Technology Madras, 
Chennai~600036, India}
\begin{abstract}
Assuming that high energy effects may alter the standard dispersion relations 
governing quantized fields, the influence of such modifications on various 
phenomena has been studied extensively in the literature.
In different contexts, it has generally been found that, while super-luminal 
dispersion relations hardly affect the standard results, sub-luminal relations 
can lead to (even substantial) modifications to the conventional results.
A polymer quantized scalar field is characterized by a series of modified
dispersion relations along with suitable changes to the standard measure 
of the density of modes.
Amongst the modified dispersion relations, one finds that the lowest in the
series can behave sub-luminally over a small domain in wavenumbers.
In this work, we study the response of a uniformly rotating Unruh-DeWitt 
detector that is coupled to a polymer quantized scalar field. 
While certain sub-luminal dispersion relations can alter the response of the 
rotating detector considerably, in the case of polymer quantization, due to 
the specific nature of the dispersion relations, the modification to the 
transition probability rate of the detector does not prove to be substantial.
We discuss the wider implications of the result.
\end{abstract}
\maketitle

\section{Introduction}

Despite decades of sustained effort, a viable quantum theory of gravity 
continues to elude us. 
In such a scenario, over the last twenty five years or so, a variety of 
approaches have been constructed {\it by hand}\/ to investigate possible 
imprints of Planck scale effects on phenomena involving matter fields 
(see, for instance, the reviews~\cite{AmelinoCamelia:2004hm,Mattingly:2005re,
Hossenfelder:2009nu,AmelinoCamelia:2008qg}). 
These methods evidently involve a new scale (often assumed to be of the 
order of the Planck scale) and they strive to capture one or more features 
that are expected to arise in the complete quantum theory of gravity.
An important goal is to utilize these approaches to examine whether high 
energy effects will modify phenomena at observably low energies.

\par

One such phenomenological approach is the approach referred to as polymer 
quantization~\cite{Hossain:2010eb}.
The method of polymer quantization can be said to be inspired by loop 
quantum gravity~\cite{Thiemann:2007zz,Rovelli:2004tv}.
(We should hasten to clarify that the approach we shall consider is different 
from another related method, also inspired by loop quantum gravity, and often 
referred to by a very similar name~\cite{Ashtekar:2002vh}. 
In this approach the configuration space is considered to be discrete, whereas 
in the approach that we shall adopt, it remains continuous.)
Using the standard Fourier decomposition of a field into oscillators and
the polymer method of quantization of the oscillators~\cite{Ashtekar:2002sn}, 
one can arrive at a modified propagator governing a quantum field in the 
Minkowski spacetime (see Ref.~\cite{Hossain:2010eb}; for a very recent 
discussion, see Ref.~\cite{Hossain:2017poa}).
While the modified propagator is identical to the conventional propagator 
(in Fourier space) at low energies, it behaves differently at high 
energies (in this context, see Ref.~\cite{Garcia-Chung:2016buw}).
It is the modified propagator which we shall utilize in this work to study 
a phenomenon closely related to the Unruh effect in flat spacetime.

\par

The Unruh effect refers to the thermal nature of the Minkowski vacuum 
when viewed by an observer in motion along a uniformly accelerated 
trajectory (for the original efforts, see Refs.~\cite{Fulling:1972md,
Unruh:1976db,DeWitt:1980hx}; for relatively recent reviews, see 
Refs.~\cite{Sriramkumar:1999nw,Crispino:2007eb,Sriramkumar:2016nmn}).
It has a close relation to Hawking radiation from black holes~\cite{Hawking:1974sw}
and, due to this reason, the effect provides an important scenario to investigate 
possible quantum gravitational effects.
In fact, the question of Unruh effect in polymer quantization has received 
some attention recently in the literature~\cite{Hossain:2014fma,Hossain:2015xqa,
Kajuri:2015oza,Husain:2015tna}. 
On the one hand, it has been claimed that, in polymer quantization, the 
Rindler vacuum may {\it not}\/ be inequivalent to the Minkowski 
vacuum~\cite{Hossain:2014fma,Hossain:2015xqa}.
On the other, it has been argued that the response of a uniformly accelerated 
Unruh-DeWitt detector coupled to a polymer quantized field would not
vanish~\cite{Kajuri:2015oza}.
In fact, it has been found that even an inertial detector will respond non-trivially 
(under certain conditions) in polymer quantization~\cite{Kajuri:2015oza,Husain:2015tna}.

\par

It is in such a situation that we choose to study the response of a rotating
Unruh-DeWitt detector that is coupled to a polymer quantized field in this
work~\cite{Letaw:1980yv,Bell:1986ir,Davies:1996ks,Unruh:1998gq,Korsbakken:2004bv,
Gutti:2010nv}.
As we shall see, the propagator in polymer quantization can be expressed as 
a series of propagators described by specific modified dispersion relations,
along with corresponding changes to the measure of the density of the 
modes~\cite{Hossain:2010eb}.
Since modified dispersion relations break Lorentz invariance, the corresponding
propagators do not prove to be time translation invariant in the frame of a
uniformly accelerated detector.
This aspect makes it rather difficult to explicitly evaluate the transition 
probability of an accelerated detector.
In contrast, since modified dispersion relations preserve rotational invariance, 
the corresponding propagators prove to be time translation invariant in the frame 
of a rotating detector, a property which allows the transition probability rate 
to be evaluated~\cite{Gutti:2010nv}.
Actually, it is such a feature that has recently been exploited to study the 
response of an inertial detector that is coupled to a polymer quantized 
field~\cite{Husain:2015tna}. 

\par

This paper is organized as follows. 
In the following section, we shall quickly review the response of inertial
and rotating Unruh-DeWitt detectors that are coupled to a massless scalar 
field governed by the standard linear dispersion relation in Minkowski
spacetime. 
In Sec.~\ref{sec:rd-mdr}, we shall briefly discuss the response of these 
detectors when they are coupled to a scalar field characterized by a modified 
dispersion relation. 
In Sec.~\ref{sec:pq}, we shall consider the response of detectors coupled 
to a scalar field described by polymer quantization.
We shall conclude with a summary of the results and their implications
in the final section.

\par

At this stage, a couple of words on our conventions and notations are in order.
We shall set $\hbar=c=1$ and, for simplicity in notation, we shall denote the 
spacetime coordinates $x^{\mu}$ as ${\tilde x}$.
We shall work in $(3+1)$-spacetime dimensions.
As far as the spatial coordinates ${\bm x}$ are concerned, we shall be working 
with either the Cartesian coordinates $(x,y,z)$ or the cylindrical polar 
coordinates $(\rho,\theta,z)$, as convenient.


\section{Detector coupled to the standard scalar field}\label{sec:rd-sc}

In this section, we shall rapidly summarize the response of inertial
and rotating Unruh-DeWitt detectors coupled to the standard, massless 
scalar field in flat spacetime.
Since the inertial and rotating trajectories are integral curves of 
timelike Killing vector fields, typically, one first evaluates the 
Wightman function along the trajectory of the detector and attempts 
to Fourier transform the resulting Wightman function.
It is well known that the inertial detector does not respond and, in
the case of the rotating detector, while the response proves to be
non-zero, the integral cannot be evaluated analytically, but can be 
easily computed numerically (see, for instance, Refs.~\cite{Letaw:1980yv,
Gutti:2010nv}).
However, it proves to be convenient to express the Wightman function as
a sum over the normal modes and evaluate the Fourier transform (with 
respect to the differential proper time) first before evaluating the sum.
In the rotating case, though the final sum needs to be calculated 
numerically, this method happens to be rather effective as the sum 
converges very quickly.
Importantly, as we shall illustrate, this method can be immediately 
extended to situations wherein the field is described by a modified 
dispersion relation~\cite{Gutti:2010nv}.


\subsection{The Unruh-DeWitt detector}

A detector can be considered to be an operational tool in an attempt 
to define the concept of a particle in a generic situation.
It corresponds to an idealized point like object, whose motion is 
described by a classical worldline, but which nevertheless possesses 
internal, quantum energy levels.
The detectors are basically described by the interaction Lagrangian 
for the coupling between the degrees of freedom of the detector and 
the quantum field.
The simplest of the different possible detectors is the monopole detector 
originally due to Unruh and DeWitt~\cite{Unruh:1976db,DeWitt:1980hx}.

\par

Consider a Unruh-DeWitt detector that is moving along a trajectory 
${\tilde x}(\tau)$, where $\tau$ is the proper time in the frame of 
the detector.
The interaction of the Unruh-DeWitt detector with a canonical, real 
scalar field~$\phi$ is described by the interaction Lagrangian
\begin{equation}
{\cal L}_{\rm int}[\phi({\tilde x})]
= {\bar c}\, m(\tau)\, \phi\left[{\tilde x}(\tau)\right],
\label{eqn:lint}
\end{equation}
where ${\bar c}$ is a small coupling constant and $m$ is the 
detector's monopole moment.
Let us assume that the quantum field ${\hat \phi}$ is in the vacuum
state, say, $\vert 0\rangle$, and the detector is in its ground state
with zero energy. 
It is then straightforward to establish that the transition probability
of the detector to be excited to an energy state with energy eigen
value $E>0$ can be expressed as 
\begin{eqnarray}
P(E) = \int\limits_{-\infty}^\infty {\rm d}\tau\, 
\int\limits_{-\infty}^\infty {\rm d}\tau'\, 
{\rm e}^{-i\, E\,(\tau-\tau')}\, 
G^{+}\left[{\tilde x}(\tau), {\tilde x}(\tau')\right],\;\;\;\;
\label{eqn:tp}
\end{eqnarray}
where $G^{+}\left[{\tilde x}(\tau), {\tilde x}(\tau')\right]$ 
is the Wightman function defined as
\begin{equation}
G^{+}\left[{\tilde x}(\tau), {\tilde x}(\tau')\right]
=\langle 0 \vert 
{\hat \phi}\left[{\tilde x}(\tau)\right]\,
{\hat \phi}\left[{\tilde x}(\tau')\right]
\vert 0 \rangle.\label{eqn:wfn}
\end{equation}

\par

When the Wightman function is invariant under time translations in the 
frame of the detector---as it can occur, for example, in cases wherein 
the detector is moving along the integral curves of timelike Killing 
vector fields---one has
\begin{equation}
G^{+}\left[{\tilde x}(\tau), {\tilde x}(\tau')\right]
= G^{+}(\tau-\tau').
\end{equation}
In such situations, the transition probability of the detector 
simplifies to
\begin{equation}
P(E)= \lim_{T\to \infty}\;\int\limits_{-T}^{T}\; \frac{{\rm d}v}{2}\, 
\int\limits_{-\infty}^{\infty}\, {\rm d}u\; 
{\rm e}^{-i\,E\, u}\; G^{+}(u),\label{eq:tp} 
\end{equation}
where 
\begin{equation}
u=\tau-\tau',\quad v=\tau+\tau'.\label{eq:uv}
\end{equation}
The above expression then allows one to define the transition probability 
rate of the detector to be
\begin{eqnarray}
R(E) = \lim_{T\to \infty}\; \frac{P(E)}{T}
= \int\limits_{-\infty}^\infty\, {\rm d}u\; 
{\rm e}^{-i\, E\, u}\; G^{+}(u).\label{eq:tpr}
\end{eqnarray}

\par

For the case of the canonical, massless scalar field, in $(3+1)$-spacetime
dimensions, the Wightman function $G^{+}\left({\tilde x}, {\tilde x}'\right)$ 
in the Minkowski vacuum is given by
\begin{equation}
G^+\left({\tilde x},{\tilde x}'\right)
=-\frac{1}{4\, \pi^2}\,
\left[\frac{1}{(t-t'-i\,\epsilon)^2
-\left({\bm x}-{\bm x}'\right)^2}\right],\label{eq:wfn-mv}
\end{equation}
where $\epsilon\to 0^{+}$ and $(t,{\bm x})$ denote the Minkowski 
coordinates.
Given a trajectory ${\tilde x}(\tau)$ that is an integral curve of a timelike
Killing vector field, the response of the detector is usually obtained by 
substituting the trajectory in this Wightman function and evaluating the 
transition probability rate~(\ref{eq:tpr}).
Instead, let us express the Wightman function as a sum over the normal
modes, evaluate the integral over $u$ first, before evaluating the sum. 


\subsection{Response of the inertial detector}

Before considering the case of the rotating detector, it is instructive to 
consider the rather simple case of an inertial detector that is moving 
with a constant velocity, say, ${\bm v}$.
The trajectory of the detector can be expressed as ${\tilde x}(\tau)
=[t(\tau),\, {\bm x}(\tau)]=(\gamma\,\tau,\,\gamma\, {\bm v}\,\tau)$, 
where $\gamma=(1-\vert {\bm v}\vert^2)^{-1/2}$ is the Lorentz factor.
The Wightman function evaluated in the Minkowski vacuum associated with the 
scalar field can be expressed as a sum over the normal modes as follows:
\begin{equation} 
G^+({\tilde x},{\tilde x}')
=\int\f{\d^3{\bm k}}{(2\,\pi)^3\, (2\,\omega)}\,
{\rm e}^{-i\,\omega\,(t-t')}\, {\rm e}^{i\,{\bm k}\cdot({\bm x}-{\bm x}')},
\label{eq:Gs-snm}
\end{equation}
where, for the massless scalar field of our interest, $\omega=\vert {\bm k}\vert
\ge 0$.
Let us now substitute the trajectory for the inertial detector in the above 
Wightman function and use the resulting expression to calculate the 
transition probability rate~(\ref{eq:tpr}).
Upon carrying out the integral over~$u$, we obtain that
\begin{equation} 
R(E)=\int\f{\d^3{\bm k}}{(2\,\pi)^2\, (2\,\omega)}\,
\delta^{(1)}\l[E+\gamma\,\l(\omega-{\bm k}\cdot{\bm v}\r)\r].
\label{eq:tpr-id}
\end{equation}
Since $E$ is positive and $(\omega-{\bm k}\cdot{\bm v})\ge 0$, the argument of the 
delta function never vanishes leading to the well known result that the inertial 
detector does not respond at all.


\subsection{Response of the rotating detector}

Let us now turn to the case of the rotating detector.
In this case, it proves to be more convenient to work with the cylindrical
polar coordinates, say, $(\rho,\theta,z)$, than the Cartesian coordinates.
It is straightforward to show that the Minkowski Wightman 
function~(\ref{eq:Gs-snm}) can be expressed in terms of the modes associated
with the cylindrical polar coordinates as follows:
\begin{eqnarray}
G^{+}({\tilde x}, {\tilde x}')
&=&\sum^{\infty}_{m=-\infty}\,
\int\limits_0^\infty\f{\d q\,q}{(2\,\pi)^2}\,
\int\limits_{-\infty}^{\infty}\f{\d k_z}{(2\,\omega)}\; 
{\rm e}^{-i\,\omega\,(t-t')}\nn\\
& &\times\,
J_{m}(q\,\rho)\, J_{m}(q\,\rho')\, {\rm e}^{i\,m\,(\theta-\theta')}\, 
{\rm e}^{i\,k_z\,(z-z')},\nn\\
\label{eq:Gs-snm-cpc}
\end{eqnarray}
where $J_m(x)$ denotes Bessel function of the first kind and of order $m$, 
while $\omega=k=\sqrt{q^2+k_z^2}\ge 0$.

\par

\begin{figure}[!t]
\begin{center}
\includegraphics[width=8.75cm]{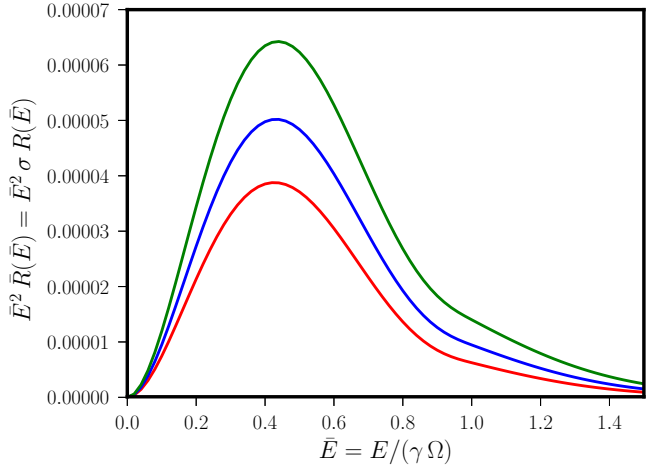}
\end{center}
\caption{The transition probability rate of the rotating Unruh-DeWitt detector 
that is coupled to a massless scalar field which is governed by the standard, 
Lorentz invariant, linear dispersion relation. 
The three curves correspond to the following choices of the dimensionless 
velocity parameter $\sigma\,\Omega$:~$0.325$ (in red), $0.350$ (in blue) and
$0.375$ (in green).}
\label{fig:rd-sc}
\end{figure}

Consider a detector moving on a circular trajectory with a radius $\sigma$ 
and angular velocity $\Omega$ in the $z=0$ plane.
The trajectory of the detector can be expressed in terms of the cylindrical
polar coordinates and the proper time as $\tx(\tau)=[t(\tau), {\bm x}(\tau)]
=(\gamma\, \tau,\,\sigma,\,\gamma\, \Omega\, \tau,\,0)$, where $\gamma
=\l[1-(\sigma\, \Omega)^2\r]^{-1/2}$ is the Lorentz factor associated with 
the trajectory.
The transition probability rate of the detector moving along the above 
trajectory can be obtained by substituting the trajectory in the 
expression~(\ref{eq:Gs-snm-cpc}) for the Wightman function and 
calculating the integral~(\ref{eq:tpr}).
We find that the resulting transition probability rate can be expressed as
\begin{eqnarray}
R(E)
&=&\sum_{m=-\infty}^{\infty}
\int\limits_0^\infty\frac{\d q\,q}{2\,\pi}\,
J_{m}^{2}(q\,\sigma)\nn\\
& &\times\,\int\limits_{-\infty}^\infty\f{\d k_z}{2\,\omega}\,
\delta^{(1)}[E+\gamma\,\l(\omega-m\,\Omega\r)].\label{eq:tpr-rd-ov}
\end{eqnarray}
This integral can be rewritten as
\begin{eqnarray}
R(E) 
&=&\sum_{m=-\infty}^{\infty}
\int\limits_0^\infty\frac{\d k\,k}{2\,\pi}\;
\delta^{(1)}[E+\gamma\,\l(k-m\,\Omega\r)]\nn\\
& &\times\,\int\limits_{0}^{\pi/2}\d \alpha\,{\rm cos}\,\alpha\,
J_{m}^{2}(k\,\sigma\, {\rm cos}\,\alpha),
\label{eq:tpr-cc}
\end{eqnarray}
where, for convenience, we have used the fact that $\omega=k$, as is appropriate 
for positive frequency modes.
The angular integral over $\alpha$ can be evaluated using the standard 
integral~\cite{Abramowitz:1972}
\begin{eqnarray}
& &\!\!\!\!\!\!\!\!\!\!\!\!\!\!\!\!\!
\int\limits_{0}^{\pi/2} \d\alpha\,
{\rm cos}\,\alpha\,J_m^2(z\,{\rm cos}\,\alpha)\nn\\
&=&\frac{z^{2\,m}}{\Gamma(2\,m+2)}\, 
{}_1F_{2}[m+1/2;m+3/2,2m+1;-z^2]\nn\\
\end{eqnarray}
for ${\rm Re}.~z\ge 0$ and ${\rm Im}.~z=0$, where ${}_1F_{2}[a;b,c;x]$
represents the generalized hypergeometric function.
Since $E$ and $\Omega$ are positive definite quantities by assumption and 
$k\ge 0$, the delta function in the expression~(\ref{eq:tpr-cc}) will be 
non-zero only when $m \ge {\bar E}$, where ${\bar E}=E/\l(\gamma\, \Omega\r)$ is a
dimensionless energy.
Therefore, the transition probability rate of the detector reduces to
\begin{eqnarray}
\!\!
{\bar R}({\bar E})
&\equiv& \sigma\, R({\bar E})\nn\\
&=&\frac{1}{2\,\pi\,\gamma}\sum^{\infty}_{m\ge {\bar E}}\,
\l[\frac{(\sigma\,k_\ast)^{2\,m+1}}{\Gamma(2\,m+2)}\r]\nn\\
& &\times\,{}_1F_2[m+1/2;m+3/2,2\,m+1;-(\sigma\,k_\ast)^2],\;\;\nn\\
\label{eq:tpr-rd} 
\end{eqnarray}
where $k_\ast= (m-{\bar E})\,\Omega$ corresponds to the value $k$ when the 
argument of the delta function in Eq.~(\ref{eq:tpr-cc}) vanishes.
It is interesting to note that, for a given energy ${\bar E}$, the detector 
seems to respond to modes whose energy $\omega~(=\!k)$ are proportional to 
the angular frequency $\Omega$ of the detector.
While it does not seem to be possible to carry out the above sum analytically, 
it converges very quickly and hence it is easy to evaluate numerically.
In Fig.~\ref{fig:rd-sc}, we have plotted the transition probability rate 
of the rotating detector (with respect to the dimensionless energy ${\bar E}$)
evaluated numerically using the above result for different values of the 
dimensionless velocity parameter $\sigma\,\Omega$.
We should clarify that we have arrived at these results by taking into account 
the contributions due to the first ten terms in the sum~(\ref{eq:tpr-rd}).
We find that the next ten terms contribute less than $0.01\%$, indicating that 
the sum indeed converges quickly and that the contributions due to the higher 
terms are insignificant.


\section{Detector coupled to a scalar field governed by a modified dispersion 
relation}\label{sec:rd-mdr}

Let us now briefly discuss the response of inertial and rotating detectors 
that are coupled to a scalar field governed by a dispersion relation, say, 
$\omega(k)$, which is no more linear.
Such dispersion relations can, for instance, arise in theories which break 
Lorentz invariance.
It can be easily shown that, in such a situation too, the Minkowski Wightman 
function can be expressed in the form~(\ref{eq:Gs-snm}) as in the standard 
case, but with the quantity $\omega(k)$ now being determined by the modified 
dispersion relation. 

\par

If the Wightman function is given by~(\ref{eq:Gs-snm}), it is then clear that 
the response of an inertial detector can also be expressed in the 
form~(\ref{eq:tpr-id}), as in the standard case.
Let us first consider a completely super-luminal dispersion relation wherein 
$\omega(k) \geq k$ for all $k$.
Since ${\bm k}\cdot{\bm v} \le k$ (as $\vert {\bm v}\vert <1$), for a 
super-luminal dispersion relation ${\bm k}\cdot{\bm v}<\omega$ and, hence, 
$\omega-{\bm k}\cdot{\bm v}>0$ for all $k$.
Therefore, the argument of the delta function in the expression~(\ref{eq:tpr-id})
never goes to zero, implying a vanishing detector response.
In contrast, consider a field governed by a sub-luminal dispersion relation 
wherein $\omega(k) < k$ over some range of $k$.
Over this domain in $k$, it is possible that $\omega-{\bm k}\cdot{\bm v} < 0$
for suitable values of the detector energy $E$ and the speed $\vert {\bm v}\vert$
of the detector.
These modes of the field can excite the detector, {\it provided the velocity of
the detector is non-zero}.\/
(In certain cases, depending on the form of the dispersion relation, there can
also arise a critical velocity, only beyond which the detector would respond, 
as we shall encounter in the following section.)
In other words, even inertial detectors with possibly a threshold velocity (which 
will, in general, depend on the internal energy $E$ of the detector) may respond 
when they are coupled to a field that is characterized by a sub-luminal dispersion 
relation.
This violation of Lorentz invariance should not come as a surprise as it is a
characteristic of fields governed by modified dispersion relations.

\par

In the standard case, the Wightman function~(\ref{eq:Gs-snm}) in the Minkowski 
vacuum could be written as~(\ref{eq:Gs-snm-cpc}) in the cylindrical polar coordinates.
In fact, this proves to be true even for the case of a scalar field described 
by a modified dispersion relation.
Hence, the transition probability rate of a rotating detector coupled to
such a field is again given by~(\ref{eq:tpr-rd-ov}), with $\omega$ [and,
later, the corresponding $k_\ast$---cf.~Eq.~(\ref{eq:tpr-rd})] suitably 
redefined.
Using these results, one can show that, while the super-luminal dispersion 
relations hardly affect the response of the rotating detector, sub-luminal
dispersion relations---depending on their shape---can substantially alter 
the response (for more details and illustration of the modified response 
in specific cases, see Ref.~\cite{Gutti:2010nv}).


\section{Detector coupled to the polymer quantized scalar 
field}\label{sec:pq}

Let us now turn to the primary case of our interest, \viz the response of 
a detector coupled to a polymer quantized scalar field.


\subsection{The Wightman function in polymer quantization}

\begin{figure}[!t]
\begin{center}
\includegraphics[width=8.75cm]{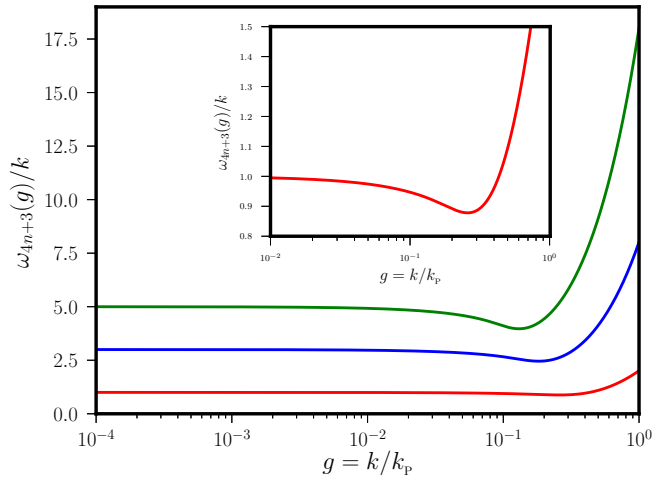}
\end{center}
\caption{The behavior of $\omega_{4n+3}(g)/k$ has been plotted 
as a function of $g=k/k_{_{\rm P}}$ for $n=0$ (in red), $n=1$ (in blue) and 
$n=2$ (in green).
The dispersion relation proves to be sub-luminal in the $n=0$ case for a small 
range of $k$ near $k_{_{\rm P}}$, while it is always super-luminal for $n>0$.
The sub-luminal behavior in the $n=0$ case is clear from the inset in the figure.}
\label{fig:omega4np3}
\end{figure}

As we had mentioned in the introductory section, the polymer quantized field can
be considered as a series of modified dispersion relations of a specific form, 
along with suitable changes to the density of modes.
In (3+1)-dimensions, the Wightman function evaluated under the polymer quantization 
procedure in the Minkowski vacuum is found to be~\cite{Hossain:2010eb, Husain:2015tna}
\begin{eqnarray}
G^+_{_{\rm P}}({\tilde x},{\tilde x}')
&=&\sum_{n=0}^{\infty}\int\frac{\d^3{\bm k}}{(2\,\pi)^3}\,
\vert c_{4n+3}(k)\vert^2\,\nn\\ 
& &\times\, {\rm e}^{-i\,\omega_{4n+3}(k)(t-t')}\, 
{\rm e}^{i\,{\bm k}\cdot({\bm x}-{\bm x}')},\label{eq:G-p}
\end{eqnarray}
where, as before, $k= \vert{\bm k}\vert$, while the quantity $\omega_{4n+3}(g)$ is 
given by
\begin{equation}
\f{\omega_{4n+3}(g)}{k_{_{\rm P}}}
=\f{g^2}{2}\;\biggl\{B_{2n+2}\l[1/\l(4\,g^2\r)\r]
-A_{0}\l[1/\l(4\,g^2\r)\r]\biggr\},
\end{equation}
with $g= k/k_{_{\rm P}}$ and $k_{_{\rm P}}$ being the polymer energy scale, 
which is usually assumed to be the Planck scale.
The quantities $A_r(x)$ and $B_r(x)$ denote the Mathieu characteristic value 
functions\footnote{We should clarify that the Mathieu characteristic value 
functions were written as $A_r(g)$ and $B_r(g)$ in the original 
work~\cite{Hossain:2010eb}.
However, in order to be consistent with the Mathieu differential equation 
describing the polymer quantized massless scalar field, they have to be 
actually written as $A_r[1/(4\,g^2)]$ and $B_r[1/(4\,g^2)]$.}.
At small $g$, one finds that $\omega_{4n+3}\simeq (2\,n+1)\, k$, which is clear
from Fig.~\ref{fig:omega4np3}, wherein we have plotted the quantity 
$\omega_{4n+3}$ as a function of $g$ for the first few values of $n$. 

\par

Moreover, the polymer coefficients $c_{4n+3}(k)$ are defined by the integral
\begin{eqnarray}
c_{4n+3}(k)
&=&\frac{i}{\pi\,\sqrt{k_{_{\rm P}}}}\,
\int\limits_0^{2\,\pi}\,\d u\, se_{2n+2}\l[1/\l(4\,g^2\r),u\r]\nn\\
& &\times\, \f{\pa\, ce_0\l[1/(4\,g^2),u\r]}{\pa u},\label{eq:c-4np3}
\end{eqnarray}
where $se_r(x,q)$ and $ce_r(x,q)$ are the elliptic sine and cosine functions, 
respectively~\cite{Abramowitz:1972}. 
It is useful to note that for $g\ll 1$, $\vert c_{4n+3}(k)\vert \simeq 
1/(\sqrt{2\,k})$, for $n=0$, which corresponds to the standard 
result~\cite{Hossain:2010eb}.

\par

In summary, three new features are encountered in polymer quantization when 
compared to the standard case.
Firstly, the quantity $\omega(k)$ in the exponential is replaced by 
$\omega_{4n+3}(k)$, in a fashion similar to that of a quantum field governed 
by a modified dispersion relation.
Secondly, the standard measure in the integral over the modes---\viz
$1/\sqrt{2\, k}$---is replaced by $c_{4n+3}(k)$.
It should be pointed out that, in the case of a field described by a modified 
dispersion relation, this measure would have been given by $1/\sqrt{2\, \omega(k)}$.
Lastly, there occurs an infinite sum over the polymer index $n$, which is an
aspect that is peculiar to polymer quantization.


\subsection{The case of the inertial detector}\label{subsec:ic}
 
Let us first revisit the response of the inertial detector in polymer 
quantization, which has been studied recently~\cite{Husain:2015tna}.

\par

\begin{figure}[!t]
\begin{center}
\includegraphics[width=8.75cm]{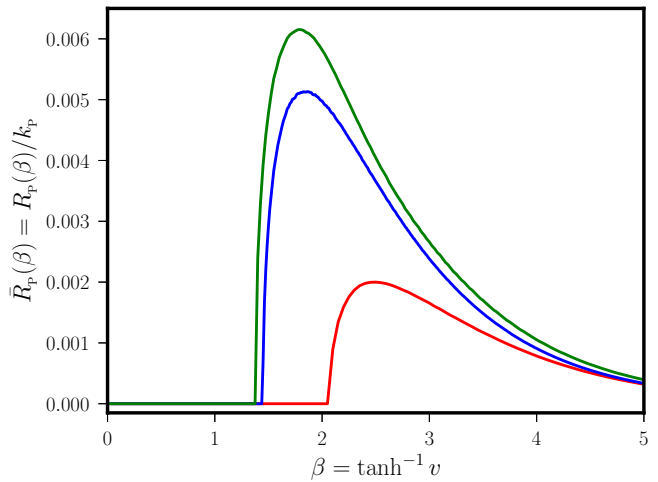}
\end{center}
\caption{The dimensionless transition probability rate ${\bar R}_{_{\rm P}}$ 
[cf.~Eq.~(\ref{eq:tpr-id-pq})] of an inertial detector that is coupled to a
polymer quantized scalar field.
We should stress that the transition probability rate has been plotted as a 
function of the rapidity parameter $\beta={\rm tanh}^{-1}\,v$ for specific 
values of ${\bar E}=E/k_{_{\rm P}}$.
Also, for reasons mentioned, we have considered only the contribution due 
to the $n=0$ case. 
The different curves correspond to ${\bar E}=0.1$ (in red), $0.01$ (in blue) 
and $0.001$ (in green).
These plots match the results that have been recently obtained in the
literature~\cite{Husain:2015tna}.
Note that, for a given ${\bar E}$, there is a threshold velocity for the 
detector to respond.
The threshold velocity seems to become smaller as ${\bar E}$ decreases.
The critical velocity beyond which an inertial detector begins to respond 
is determined by the condition that the argument of the $\Theta$ function 
in Eq.~(\ref{eq:tpr-id-pq}) turns positive.
The associated critical rapidity parameter is determined by the relation  
${\rm tanh}\,\beta_{\rm c}={\rm Min}.~[\omega_{3}(k)/g]$, which leads to 
$\beta_{\rm c}\simeq 1.3267$.}
\label{fig:rid-pq1}
\end{figure}

\begin{figure}[!t]
\begin{center}
\includegraphics[width=8.75cm]{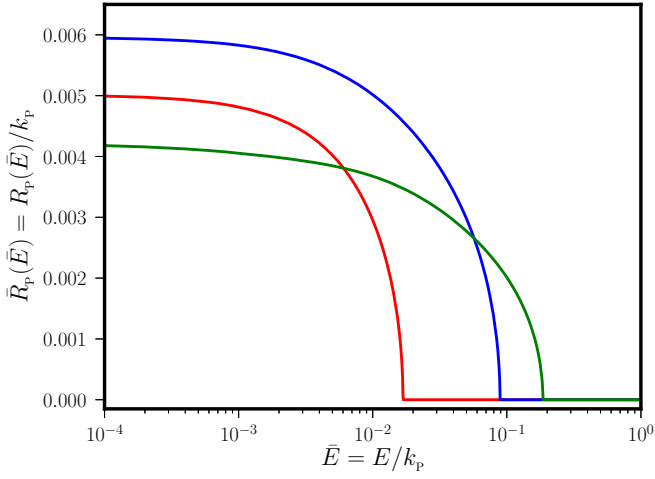}
\end{center}
\caption{The dimensionless transition probability rate ${\bar R}_{_{\rm P}}$ 
of an inertial detector that is coupled to the polymer quantized scalar field
has been plotted as a function of the dimensionless energy ${\bar E}=E/k_{_{\rm P}}$
for different values of $\beta$, which corresponds to different velocities of 
the detector.
The different curves correspond to $\beta=1.5$ (in red), $2.0$ (in blue) and
$2.5$ (in green).}
\label{fig:rid-pq2}
\end{figure}

In such a case, upon considering the Wightman function~(\ref{eq:G-p}) along
the inertial trajectory ${\tilde x}(\tau)=(\gamma\,\tau,\gamma\, {\bm v}\,
\tau)$, where $\gamma=(1-\vert {\bm v}\vert^2)^{-1/2}$ and calculating the
corresponding transition probability rate, we obtain that
\begin{eqnarray}
{\bar R}_{_{\rm P}}({\bar E})
&=& \f{R_{_{\rm P}}({\bar E})}{k_{_{\rm P}}}\nn\\
&=& \frac{1}{2\,\pi}\, \sum_{n=0}^{\infty}
\int\limits_{-⁠\infty}^{\infty}\d^{2} {\bm k}_\perp\,
\int\limits_{0}^{\infty}\d k_\parallel\, k_\parallel\, \vert c_{4n+3}(k)\vert^2\nn\\
& &\times\,\delta^{(1)}\l[E+\gamma\, \omega_{4n+3}(k)-\gamma\, k_\parallel\, v\r],
\end{eqnarray}
where ${\bar E}=E/⁠k_{_{\rm P}}$ and $v=\vert {\bm v}\vert$, while $k_{\parallel}$ 
and ${\bm k}_\perp$ denote the components of ${\bm k}$ that are parallel and 
perpendicular to the velocity vector ${\bm v}$.
Upon making the change of variables to $k_\parallel=k\,{\rm cos}\,\theta$ and 
${\bm k}_\perp=k\,{\rm sin}\,\theta$, we obtain that 
\begin{eqnarray}
{\bar R}_{_{\rm P}}({\bar E})
&=& \frac{1}{2\,\pi\,\gamma\,v}\,
\sum_{n=0}^{\infty}
\int\limits_{0}^{\infty}\d k\, k\, \vert c_{4n+3}(k)\vert^2\nn\\
& &\times\,\int\limits_{-⁠1}^{1}\d\, ({\rm cos}\,\theta)\, 
\delta^{(1)}\l[{\rm cos}\,\theta
-\f{E+\gamma\, \omega_{4n+3}(k)}{\gamma\, k\, v}\r].\nn\\
\end{eqnarray}
Note that the above integral is non-⁠zero, only if
\begin{equation}
\vert E+\gamma\, \omega_{4n+3}(k)\vert<\gamma\, k\, v,
\end{equation}
which leads to the following expression for the transition probability rate:
\begin{eqnarray}
{\bar R}_{_{\rm P}}({\bar E})&=& \frac{1}{2\,\pi\,\gamma\,v}\,
\sum_{n=0}^{\infty}
\int\limits_{0}^{\infty}\d k\, k\, \vert c_{4n+3}(k)\vert^2\nn\\
& &\times\,\Theta\bigl[\gamma\, k\, v-\vert E+\gamma\, 
\omega_{4n+3}(k)\vert\bigr],\label{eq:tpr-id-pq}
\end{eqnarray}
where $\Theta(x)$ denotes the theta function.
It seems difficult to evaluate the above transition probability rate  
analytically.
Hence, we have to resort to numerics~\cite{Husain:2015tna}.
We shall first need to determine the domain in $k$ (or, equivalently, $g$) over 
which the $\Theta$ function contributes.
It is expected to contribute when $\omega_{4n+3}(k)$ behaves sub-luminally.
It is clear from the plots in Fig.~\ref{fig:omega4np3} that the function  
$\omega_{4n+3}(k)$ is always super-luminal when $n>0$.
Therefore, these terms are not expected to contribute to the response of the 
detector.
Moreover, in the $n=0$ case, the sub-luminal behavior occurs roughly over the
small domain wherein $0.01\lesssim g\lesssim 1$.
It is these modes which we need to integrate over.
We evaluate the quantity $c_{4n+3}(k)$ using the definition~(\ref{eq:c-4np3})
before going on to carry out the integral over the relevant domain in $k$ 
(determined by the $\Theta$ function) to arrive at the transition probability 
rate of the detector.
We find that, because the integrand in Eq.~(\ref{eq:c-4np3}) is well behaved, 
both the integrals can be evaluated with even the simplest of methods. 
We make use of the Simpson's rule to carry out these integrals.
We should emphasize that we have checked the accuracy of the integrations
involved by working with a larger number of steps as well as using the  
more accurate Bode's rule.
We find that the integrations we have carried out are accurate to better 
than $0.01\%$.
In Fig.~\ref{fig:rid-pq1}, we have plotted the dimensionless transition 
probability rate as a function of the rapidity parameter~$\beta
={\rm tanh}^{-1}\,v$.
These curves match the results obtained earlier~\cite{Husain:2015tna}.
In Fig.~\ref{fig:rid-pq2}, we have plotted the transition probability rate
as a function of the dimensionless energy ${\bar E}=E/k_{_{\rm P}}$ for a
few different values of $\beta$.
These results confirm the correctness of our numerical procedures.


\subsection{The case of the rotating detector}\label{subsec:rc}

To determine the response of the rotating detector coupled to a polymer
quantized field, we shall follow the same strategy that we had adopted 
earlier.
It is straightforward to establish that, when working in the cylindrical
polar coordinates, along the rotating trajectory that we had considered 
earlier, the polymer quantized Wightman function~(\ref{eq:G-p}) is given 
by
\begin{eqnarray}
G^+_{_{\rm P}}(u)
&= &\sum_{n=0}^{\infty}\sum_{m=-\infty}^{\infty} 
\int\limits_0^{\infty} \f{\d q\, q}{2\,\pi}\,
\int\limits_{-\infty}^{\infty}\f{\d k_z}{2\,\pi}\, \vert c_{4n+3}(k)\vert^2\nn\\ 
& &\,\times\, J_m^2(q\,\sigma)\, 
{\rm e}^{-i\,\l[\omega_{4n+3}(k)-\gamma\, m\, \Omega\r]\,u},
\end{eqnarray}
where, as before, $k=\sqrt{q^2+k_z^2}$.
We can convert the integrals over $q$ and $k_z$ into integrals over $k$
and a suitable angle $\alpha$, as in the standard case.
Upon doing so and carrying out the integrals over $\alpha$ as well as $u$, 
we find that the transition probability rate of the rotating detector can 
be expressed as
\begin{eqnarray}
{\bar R}_{_{\rm P}}({\bar E})
&\equiv& \sigma\, R_{_{\rm P}}({\bar E})\nn\\
&=& \sum_{n=0}^{\infty}\,
\sum^{\infty}_{m=-\infty} 
\int\limits_0^\infty \f{\d k}{2\,\pi}\; 
\f{\l(\sigma\,k\r)^{2\,m+1}}{\Gamma(2\,m+2)}\nn\\
& &\times\,2\,k\,\vert c_{4n+3}(k)\vert^2\,\nn\\
& &\times\, {}_1F_2[m+1/2;m+3/2,2\,m+1;-(\sigma\,k)^2]\nonumber \\
& &\times\, 
\delta^{(1)}\l[E+\gamma\, \omega_{4n+3}(k)-\gamma\, m\,\Omega\r].
\end{eqnarray}
The integral over $k$ can be evaluated immediately to arrive at 
\begin{eqnarray}
{\bar R}_{_{\rm P}}({\bar E})
&=&\f{1}{2\,\pi\,\gamma}\, \sum_{n=0}^{\infty}\,
\sum^{\infty}_{m\ge {\bar E}}  
\f{\l(\sigma\,k_\ast\r)^{2\,m+1}}{\Gamma(2\,m+2)}\nn\\
& &\times\,\l[\f{2\,k_\ast\,\vert c_{4n+3}(k_\ast)\vert^2}{\vert\d 
\omega_{4n+3}/\d k\vert_{k=k_\ast}}\r]\nn\\
& &\times\, {}_1F_2[m+1/2;m+3/2,2\,m+1;-(\sigma\,k_\ast)^2],\nn\\
\label{eq:tpr-kp}
\end{eqnarray}
where $k_\ast$ now denote the roots of the equation 
\begin{equation}
\omega_{4n+3}(k_\ast)=\l(m-{\bar E}\r)\,\Omega.
\end{equation}
Since $\omega_{4n+3}(k)$ is a positive definite quantity, we need to 
confine ourselves to $m\ge {\bar E}$ in the above sum, exactly as in 
the standard case.
Note that, in the standard situation, we have just the $n=0$ case, with
$\omega(k)=k$, leading to $k_\ast=\l(m-{\bar E}\r)\,\Omega$.
Also, in such a case, the quantity within the large square brackets in 
the above expression reduces to unity, thereby simplifying to the 
result~(\ref{eq:tpr-rd}) we had obtained earlier.

\par

\begin{figure}[!t]
\begin{center}
\includegraphics[width=8.75cm]{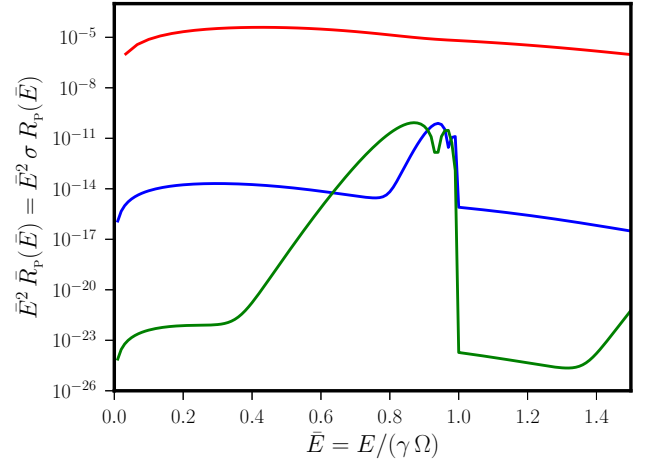}
\end{center}
\caption{The transition probability rate of the rotating detector when it 
is coupled to a polymer quantized scalar field.
The three curves correspond to $n=0$ (in red), $n=1$ (in blue) and $n=2$
(in green).
We have set $\sigma\,\Omega=0.325$ and ${\bar k}_{_{\rm P}}=\sigma\, 
k_{_{\rm P}}=10^2$ in plotting these curves.
Note that, in order to illustrate the relative magnitude of the three terms, 
in contrast to Fig.~\ref{fig:rd-sc}, we have plotted the $y$-axis on a  
logarithmic scale.
Clearly, the $n=0$ term dominates the contributions to the transition 
probability rate of the rotating detector.
Therefore, the higher order terms can be safely ignored.}
\label{fig:tpr-rd-pq-rc}
\end{figure}

In order to determine the transition probability rate of the rotating
detector, we need to first determine the roots $k_\ast$ and evaluate 
the quantities $\vert c_{4n+3}(k)\vert^2$ and $\vert\d 
\omega_{4n+3}(k)/\d k\vert$ at these $k_\ast$.
As we had mentioned in the inertial case, these seem impossible to evaluate 
analytically. 
However, we find that they can be determined numerically without much 
difficulty.
Having determined the roots $k_\ast$, the quantity $\vert\d \omega_{4n+3}(k)
/\d k\vert$ is easy to obtain.
We evaluate $c_{4n+3}(k)$ just as in the inertial case, using the Simpson's
rule. 
Once all these quantities are in hand, we also need to sum over $n$ and 
$m$ to arrive at the complete transition probability rate of the detector.
The sum over $m$ converges rapidly as in the standard case. 
In Fig.~\ref{fig:tpr-rd-pq-rc}, we have plotted the contributions due to the 
the first three terms in the sum over $n$ for specific values of the
parameters involved.
It is evident from the figure that the $n=0$ term dominates the contribution.

\par

\begin{figure}[!t]
\begin{center}
\includegraphics[width=8.75cm]{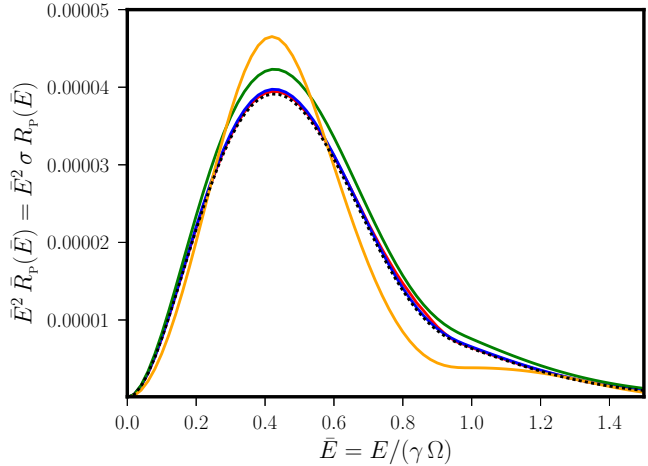}
\end{center}
\caption{The transition probability rate of the rotating detector that is
coupled to a polymer quantized field has been plotted for different 
values of $k_{_{\rm P}}$.
We have set $\sigma\,\Omega=0.325$ and have taken into consideration only
the $n=0$ contribution to the response of the detector.
Note that the different solid curves correspond to the following values of
${\bar k}_{_{\rm P}}=\sigma\, k_{_{\rm P}}$: $10^3$ (in red), $10^2$ (in blue), 
$10$ (in green) and unity (in orange).
The dotted black curve corresponds to the standard case we had plotted in
Fig.~\ref{fig:rd-sc}.
Evidently, the larger the ${\bar k}_{_{\rm P}}$, the smaller is the deviation
from the standard case.
This indicates that the high energy modifications do not alter the response of 
the rotating detector considerably.}
\label{fig:rd-pq-vkp}
\end{figure}

Let us now turn to examine if polymer quantization modifies the transition
probability rate of the rotating detector.
In Fig.~\ref{fig:rd-pq-vkp}, we have plotted the transition probability rate
of the detector for a few different values of $k_{_{\rm P}}$.
We should mention here that, as in the standard case, we have taken into 
account only the first ten contributions in the sum over $m$ in~\ref{eq:tpr-kp}
(for the $n=0$ case, as discussed above).
We have also examined and confirmed that the contributions due to the higher
terms are indeed completely insignificant.
It is clear that, even for an extreme value of ${\bar k}_{_{\rm P}}=\sigma\,
k_{_{\rm P}}=1$, the detector response does not differ considerably from the
standard case.
This suggests that polymer quantization does not alter the standard results
appreciably.


\section{Summary}

The approach due to polymer quantization takes into account certain aspects 
that are expected to arise in a plausible quantum theory of gravitation and 
arrives at a modified version of the standard Minkowski propagator. 
The response of the so-called detectors that are coupled to a scalar field 
are determined by the Fourier transform of the Wightman function governing 
the field.  
In this work, using the propagator arrived at by polymer quantization, we have 
investigated the effects of high energy physics on a variant of the Unruh effect.

\par

It is well known that, while inertial detectors do not respond in the Minkowski 
vacuum (when coupled to the standard quantum field), rotating detectors exhibit 
a non-zero response.
But, it proves to be difficult to calculate the response of the rotating detector
analytically and one needs to resort to numerics to evaluate the transition 
probability rate of the detector.
These two results are easy to understand. 
As the standard Wightman function in the Minkowski vacuum is Lorentz invariant, 
it is not surprising that inertial detectors do not respond in such a situation.
In contrast, it seems natural to expect that detectors in non-inertial motion
will, in general, respond non-trivially in the Minkowski vacuum (in this context, 
see, for instance, Ref.~\cite{Sriramkumar:1999nw}).
In this work, we have studied the response of detectors that are coupled to a 
scalar field which is quantized through the method of polymer quantization.
After revisiting the case of the inertial detector which has been studied recently, 
we had investigated the response of a rotating detector. 
It has been shown earlier that detectors which are coupled to a quantum field
that is described by super-luminal dispersion relations are hardly affected.
Also, it is known that the response of the detectors can be altered considerably
if they are coupled to a field characterized by sub-luminal dispersion relations.
In the case of a polymer quantized field, one of the dispersion relations governing 
the field behaves sub-luminally over a limited domain in wavenumber.
It is this behavior that is expected to alter the response of the rotating
detector~\cite{Gutti:2010nv}.
However, in polymer quantization, since the sub-luminal modification to the dispersion 
relation is small, we find that, the corresponding change in the response of the 
detector is also not considerable.
Our results confirm similar conclusions concerning the sub-luminal and super-luminal
dispersion relations that have been arrived at earlier.
Specifically, two phenomena where the effects due to trans-Planckian physics 
have been investigated to a considerable extent are Hawking radiation from 
black holes and the inflationary perturbation spectra. 
In both these cases, it has been found that, while super-luminal dispersion
relations hardly affect the conventional results, sub-luminal relations can,
in principle, alter (depending on the details of the dispersion relation) 
the standard results to a good extent (in this context, see the 
reviews~\cite{Jacobson:1999zk,Brandenberger:2002hs}).

\par

A couple of additional points need to be clarified concerning the responses of the 
inertial and rotating detectors that are coupled to a polymer quantized field.
While the response of an inertial detector that is coupled to the standard quantum 
field vanishes identically, the detector coupled to a polymer quantized exhibits 
a non-zero response.
This may suggest that the modifications to the response of the inertial detector
(when coupled to the polymer quantized field) are significant.
In contrast, the response of the rotating detector coupled to a polymer quantized 
field seems hardly different from the standard case.
We believe that the changes are not necessarily significant in the inertial case,
as it should be noticed that the transition probability rate in Fig.~\ref{fig:rid-pq2}
has been plotted in units of $k_{_{\rm P}}$.
In fact, it is also easy to illustrate a similar point with the rotating detector.
Note that there exists a static limit in the rotating frame.
It has been shown that a rotating detector coupled to the standard field ceases
to respond when one imposes a boundary condition on the field at the static
limit~\cite{Davies:1996ks}.
However, it is easy to argue that a rotating detector coupled to a polymer quantized
field will respond non-trivially (due to the sub-luminal nature of the dispersion
relation) in the same situation~\cite{Gutti:2010nv}.
This may again naively indicate, as in the inertial case, that the modification
is substantial.
However, we know that the modifications are only minimal in the case without the
boundary.
These arguments support the fact that the changes to the detector response in the
inertial case cannot be considered to be substantial.
Such a conclusion would also be consistent with the conclusion we have drawn in the
rotating case.

\par

Needless to add, it will be interesting to evaluate the response of a uniformly
accelerated detector that is coupled to a polymer quantized field.
However, as we had pointed out in the introductory section, the polymer Wightman 
function does not prove to be translation invariant in terms of the proper time 
in the frame of the accelerated detector.
This poses difficulty in evaluating the corresponding transition probability
rate of the detector.
One possible way to deal with this problem is to evaluate the response of the 
detector for a finite proper time interval and examine the behavior of the 
response when the duration for which the detector is kept switched on is much 
larger than the time scale associated with the 
acceleration~\cite{Sriramkumar:1994pb}.
We are currently investigating this issue.


\end{document}